\documentstyle[epsf]{mn}
\def\mmm{(m-M)$_0$}
\def\ebv{E($B-V$)~}

\def\bv{$B-V$}

\def\msun{M$_{\odot}$}
\def\gsim{\;\lower.6ex\hbox{$\sim$}\kern-7.75pt\raise.65ex\hbox{$>$}\;}
\def\lsim{\;\lower.6ex\hbox{$\sim$}\kern-7.75pt\raise.65ex\hbox{$<$}\;}

\title[NGC2660]{The intermediate age open cluster NGC 2660\thanks{
 Based on observations made in La Silla, ESO}}

\author[Sandrelli et al.]{S. Sandrelli$^{1,2}$, A. Bragaglia$^1$, M.
        Tosi$^1$, G. Marconi$^3$    \\
 $^1$ Osservatorio Astronomico di Bologna, Via Ranzani 1, I-40127 Bologna,
      Italy, 
      e-mail sandrelli, angela, tosi @bo.astro.it \\
 $^2$ Osservatorio Astronomico di Brera, Via Brera 28, I-20121 Milano,
      Italy, 
      e-mail stefano@brera.mi.astro.it \\
 $^3$ Osservatorio Astronomico Roma, Via dell'Osservatorio 5, I-00040 Monte 
      Porzio, Italy,
      e-mail marconi@coma.mporzio.astro.it}
\date{}

\begin{document}
\maketitle

\begin{abstract}
We present CCD UBVI photometry of the intermediate old open cluster NGC2660,
covering from the red giants region to about seven magnitudes below the main
sequence turn-off.

Using the synthetic Colour - Magnitude Diagram method, we estimate in a
self-consistent way values for distance modulus (\mmm $\simeq$ 12.2),
reddening (\ebv $\simeq$ 0.40), metallicity ([Fe/H] about solar), and age
($\tau$ $\lsim$ 1 Gyr). A 30$\%$ population of binary stars turns out to be
probably present.

\end{abstract}


\begin{figure*}
\vspace{11cm}
\includegraphics{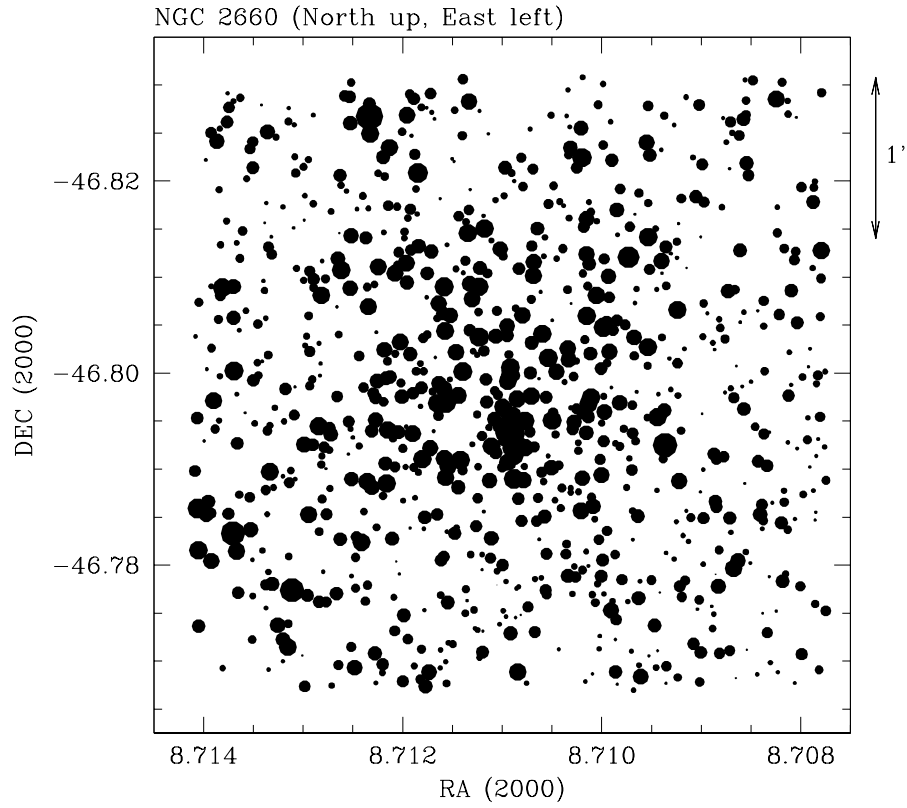}
\caption{Map of the observed field derived from the photometry of one of the
deep V frames: the bright carbon star, and a few saturated field stars are
missing. North is up and East left; the field is 3.8x3.8 arcmin$^2$.}
\label{fig-map}
\end{figure*}

\begin{figure*}
\vspace{9cm}
\includegraphics{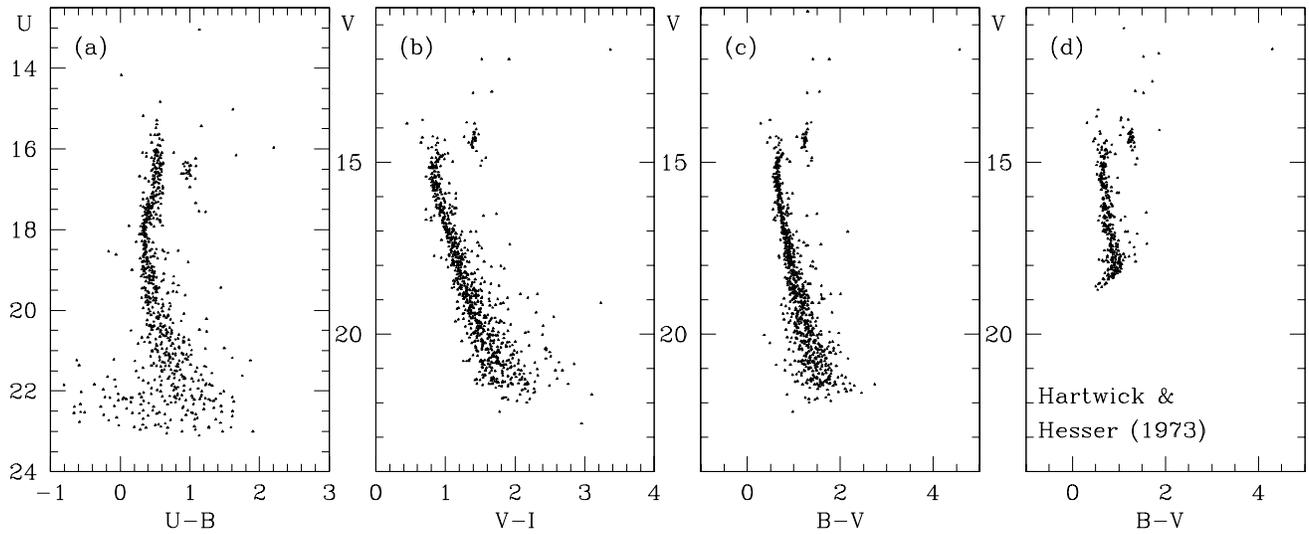}
\caption{(a,b,c) Colour Magnitude Diagrams for our sample, and (d) comparison 
with HH73}
\label{fig-cmd}
\end{figure*}

\section{Introduction}

Galactic open clusters cover a large range of distances, metallicities and
ages, which allows us to map out the disc of our Galaxy, so that they reveal
themselves to be a precious tool in the study of the chemical and dynamical
evolution of the Milky Way.

We have undertaken a program to carefully analyse open clusters 
at various positions, metallicities and ages. Our homogeneus analyses
are addressed mostly to the older objects (i.e. with ages larger than 
the Hyades). 

Distance, age, reddening and approximate metallicity of the clusters are
derived from comparison of the observed colour-magnitude diagrams (CMDs) to
synthetic ones generated by a numerical code based on stellar evolution
tracks and taking into account theoretical and observational uncertainties
(Tosi et al. 1991). This method, alternative to the classical isochrone
fitting, has proven much more powerful and successful in studying the
evolutionary status and properties of the analysed objects.
We have already applied it to several Galactic open clusters, namely:
NGC7790 (Romeo et al. 1989), NGC2243 (Bonifazi et al. 1990), 
Cr261 (Gozzoli et al. 1996), NGC6253 (Bragaglia et al. 1997), 
NGC2506 (Marconi et al. 1997), and Berkeley 21 (Tosi et al. 1998).

NGC2660 is an intermediate age open cluster located at coordinates RA(2000) =
8:42:18, DEC(2000) = --47:09, and l$_{\rm II}$ =266$^{\circ}$, b$_{\rm II}$ =
$-3.0^{\circ}$.
In the past it has also been studied because of a carbon star possibly
associated to it (GV Vel, e.g. Hartwick \& Hesser 1973, and recently
Groenewegen, van den Hoeck \& de Jong 1995).

The cluster CMD has already been published by Hartwick \& Hesser (1973,
hereafter HH73), and by Frandsen, Dreyer \& Kjeldsen (1989, hereafter FDK),
but our new data reach fainter and/or have higher photometric precision.
HH73 employed photoelectric $U, B, V$, photographic $B, V$, and photoelectric
$uvbyH_\beta$ measures to determine the following properties: \ebv $\simeq$
0.38, \mmm = 12.3$\pm0.3$, age $\sim$ 1.2 Gyr, metallicity similar to the
Hyades, and high probability of membership for the N-type carbon star.
FDK studied several open clusters to find trace of stellar activity and
variability, using repeated Johnson $V$ and Gunn $r$ photometry on a period of
a few days (4 to 12, depending on the cluster). In NGC2660 they found many
candidate variables 
(about 10\% of the stars measured on a 2$\times$3 arcmin$^2$ field). They 
argue that the cluster has a normal number of $\delta$-Scuti variables, and 
many low amplitude variables, both on the main sequence (MS) and in
the giant region. They present $V$ and $r$ light curves for three objects, one
of which looks unquestionably a variable (defined as a rotating spotted K-type
star), and two $may$ be badly sampled eclipsing variables.

The cluster metallicity has been measured by various authors, but with
inconsistent results. HH73 estimated the metallicity to be $\leq$
[Fe/H]$_{Hyades}$ on the basis of the ultraviolet excess in the two-colour
diagram of four red giant stars. Hesser \& Smith (1987), employing DDO
photometry of 13 giant stars, determined a metallicity 0.6 dex lower than in 
the Hyades, or [Fe/H]=--0.4, on a scale where [Fe/H]$_{Hyades}$=+0.2.
Geisler, Clari\'a \& Minniti (1992), using Washington photometry on cluster
giants, investigated the metallicity of several open clusters with an
accuracy of about 0.2 dex. While in most cases their metallicities show good
agreement with other determinations, NGC2660 is an exception, with
[Fe/H]=--1.05$\pm$0.16. They also claim that the Houdashelt, Frogel \& Cohen
(1992) IR observations of the cluster go in the same direction, resulting in
a CMD indicative of a quite metal-poor population. Friel (1995) cites for
NGC2660 a value of [Fe/H]=+0.06, taken from Lyng{\aa} (1987); there are no
spectroscopic measures for this cluster made by her group. The last
determination comes again from DDO photometry, but with a new abundance
calibration, and is found in Piatti, Clari\'a \& Geisler (1995): they obtain
[Fe/H]=--0.27$\pm$0.13 on 5 stars with membership information.

On the contrary, the reddening values have always been found in good
agreement. HH73 give \ebv=0.38$\pm$0.05, from a variety of methods,
comprising the position of MS stars in the two-colour diagram,  a few nearby
Cepheids, and Str\"omgren colours.
Hesser \& Smith (1987) infer a value of \ebv=0.35$\pm$0.03 from their DDO
photometry. 
Lewis \& Freeman (1989) compute, from their reddening model based on
observations of old disc giants, \ebv=0.38.

As to the age of NGC2660, HH73 determined a value of $\sim$ 1.2 Gyr using
isochrones computed by them on purpose. Lyng{\aa} (1987) cites 1.6 Gyr. 
Janes \& Phelps (1994), on the basis of the $\delta$V method,
compute an age of 0.9 Gyr, while Carraro \& Chiosi (1994) and Carraro,
Ng \& Portinari (1998) derive 0.7 Gyr (adopting [Fe/H]=+0.06).

In Section 2 we describe the observations and data analysis; in Section 3 we
present the derived CMDs involving $UBVI$ photometry and discuss the presence
of binary stars. In Section 4 we compare observed and synthetic CMDs and
derive metallicity, age, distance and reddening. Finally, the conclusions
will be reviewed in Section 5.

\section{Observation and data reductions}

NGC2660 was observed at the 0.91m Dutch telescope located in La Silla, Chile,
on May 8, 1997; the field was centered on the cluster, at coordinates
RA(2000)=8:42:36, DEC(2000)=--47:13:25.
The direct camera mounted  the CCD \#33, a Tek 512$\times$512, with a scale of
0.442 arcsec/pix, yielding a field of view of 3.8$\times$3.8 arcmin$^2$.
The observed region is shown in Fig.~\ref{fig-map}, derived from our
photometry (bright objects, like the carbon star, and a few field stars, are
missing), and oriented with North up and East left.

Observations were done in the Bessel $U, B, V$ filters (ESO \# 634, 419, 420)
and Gunn $i$ (ESO \# 465). Exposures in $B, V, i$ were both short (30 or 60
seconds) and long (10 or 15 minutes). The single 30 minutes $U$ exposure
doesn't reach as faint as the other bands, but is useful for reddening
determination, true cluster members selection and binary stars detection.

The observations were performed on a photometric night, with standard stars
fields observed before and after the cluster. The airmass was always less than
1.5, and seeing conditions were quite poor, but this did not pose any problem 
even in the (relatively) crowded central region: no star was lost due to
confusion with a nearby object. The seeing values for the four images used as 
master frames for the reduction  are indicated in Table 1.

Standard CCD reductions of bias and dark current subtraction, trimming, and
flatfield correction were performed. 

The sum of two deep frames (300 and 900 seconds) in $V$ has been used to
search for stellar objects, setting  the minimum photometric threshold for
object detection at 3$\sigma$ above the local sky background. The stars which
seemed to be saturated were recovered on a V frame of shorter exposure (30
seconds). All the identified objects were then fitted in all the others
frames.

We applied to all frames the usual procedure  for PSF study and fitting 
available in DAOPHOT--II (Stetson 1992) in IRAF\footnote{
IRAF is distributed by the NOAO, which are operated by AURA, under contract 
with NSF} environment. 

For every filter, we identified a reference frame which we used to calibrate
the PSF derived magnitudes of any other images. We chose 15-20 isolated
stars in all the four reference images and computed the aperture corrections
to the PSF estimates. In a given filter, the measured magnitude $m$ is
obtained as
$$ m = m_{psf} + A - k\cdot F_z $$
where $m_{psf}$ is the PSF derived magnitudes, $A$ the aperture correction,
$k$ the extinction coefficient for that filter and $F_z$ the airmass
(Table 1).

\begin{table}
\begin{center}
\caption{Aperture correction, airmass and extinction coefficients for the
instrumental magnitudes of the four reference frames in the $U, B, V$ and $I$
filters. The fifth column gives the corresponding seeing values.}
\begin{tabular}{c c c c c}
\hline\hline
Filter  & aperture & airmass & extinction     &seeing\\     
        &   (mag)  &         & ($k_\lambda$)  &(arcsec)\\     
\hline
$U$ 	& -0.233    &  1.17    & 0.59 & 1.64\\     
$B$ 	& -0.252    &  1.43    & 0.22 & 2.00\\     
$V$ 	& -0.315    &  1.11    & 0.12 & 1.30\\     
$I$ 	& -0.361    &  1.21    & 0.06 & 1.28\\     
\hline
\end{tabular}
\end{center}
\label{tab-corre}
\end{table}

\subsection{Photometric calibrations}

The conversion from instrumental magnitudes to the Johnson-Cousins
standard system was obtained using the following set of primary calibrators:
PG0918+019, PG1323--086, PG1525--071, Mark A (Landolt 1992).
The adopted stars span a wide range in colour (--0.27 $\leq  B-V \leq$ 1.13), 
and cover almost the whole interval of interest for this cluster, 
from the MS Turn-Off (TO) point to (with a small extrapolation) the red clump.
Standard stars fields were analysed using aperture photometry. The
calibration equations were derived using La Silla mean extinction
coefficients for the month of May, taken from the database maintained by the
the Geneva Observatory photometric group.

We obtained equations in the form:

\[ U(u,u-b) = u + 0.039(\pm 0.003)\cdot (u-b) - 5.235(\pm 0.01 ) \]

\[ B(b,b-v) = b + 0.069 (\pm 0.001) \cdot (b-v) - 3.347 (\pm 0.001) \]

\[ V(v,b-v) = v +  0.008 (\pm 0.010) \cdot (b-v) - 2.952 (\pm 0.02) \]

\[ V(v,v-i) = v +  0.098 (\pm 0.010) \cdot (v-i) - 2.946 (\pm 0.001 ) \]

\[ I(i,v-i) = i -  0.043 (\pm 0.005 ) \cdot (v-i) - 4.075 (\pm 0.001) \]

\noindent
where $u, b, v, i$ are instrumental magnitudes, and $U, B, V, I$ are the 
corresponding Johnson-Cousins magnitudes.
   
\begin{table}
\begin{center}
\caption{Completeness of our measurements. Each value is the average of 10 to
30 trials for $U, B, V$, and $I$.}
\begin{tabular}{c r r r r}
\hline\hline
Mag interval &\%U &\%B &\%V &\%I\\
\hline
13.0 &	    &      &	  & 100 \\
13.5 &	    &      &	  &  94 \\ 
14.0 &	    & 100  &      &  94 \\
14.5 &	    &  90  & 100  &  94 \\
15.0 &      &  98  &  98  &  83 \\
15.5 &      &  90  &  98  &  87 \\
16.0 & 100  &  97  & 100  &  85 \\
16.5 &  96  &  82  &  96  &  78 \\
17.0 &  98  & 100  &  98  &  71 \\
17.5 & 100  &  95  &  88  &  64 \\
18.0 &  97  &  97  &  95  &  57 \\
18.5 & 100  &  92  &  91  &  40 \\
19.0 &  95  &  93  &  87  &  10 \\
19.5 &  96  &  90  &  84  &  20 \\
20.0 &  92  &  80  &  76  & $<10$\\
20.5 &  92  &  70  &  79  &  \\
21.0 &  78  &  61  &  64  &  \\
21.5 &  74  &  47  &  52  &  \\
22.0 &  52  &  45  &  37  &  \\
22.5 &  31  &  22  &  10  &  \\
23.0 &  20  &  10  & $<10$&  \\
23.5 & $<10$& $<10$&      &  \\     
\hline
\end{tabular}
\end{center}
\label{tab-compl}
\end{table}

\subsection{Completeness analysis}

A suitable new IRAF task (ALLBIN) was prepared to test automatically the
completeness of our luminosity function in the $U, B, V$ and $I$ band. In
short, ALLBIN uses the routine ADDSTAR (in DAOPHOT--II) to add to the
original deepest frames in each filter a pattern of artificial stars,
distributed in colour as the real ones ($\simeq$ $10\%$ of the total in each
magnitude bin), at random positions. The obtained ``artificial frames'' were
then reduced using exactly the same procedure and the same PSF used for the
original images.

We considered as ``recovered'' only those stars found in their given position
and magnitude bin. The completeness was then derived as the ratio
$N_{recovered}/N_{added}$ of the artificial stars generated.

We performed 10 to 30 trials per bin in every band, depending on the 
stability of the results. The final averaged results are reported 
in Table 2.

\section{The colour-magnitude diagrams}

The stars measured in our field are 866; the corresponding CMDs
are shown in Fig.~\ref{fig-cmd}, together with the older data by HH73. The
cluster MS is well delineated, and can be followed for about
7 magnitudes fainter than the TO. There is a well populated clump
of red stars, which is attributable to the core-He burning phase, and a few
red giants are also visible. 

The carbon star, \#188 in our numbering system, is visible
at $U=19.48$, $B=16.32$, $V=11.72$, $I=8.36$; for a comparison, HH73 (in their
numbering system the star is \#9009) give $U=19.59$, $B=15.97$, and $V=11.68$.
Since the star is variable, and so bright as to be near the saturation limit,
we consider the magnitudes as acceptably similar. Taking into account
the best values for distance modulus and reddening based on the synthetic
colour-magnitude diagram method (Section 4), the absolute
magnitude of the carbon star is M$_{\rm V}$=--1.90 according to the 
Padova and FRANEC solar tracks and adopting \mmm=12.3 and \ebv=0.41, and
in the range --1.96$\leq$ M$_{\rm V} \leq$--1.55 according to the FST-ov
models, where the higher uncertainty, due to the lack of He-burning
phases, reflects in a range of equally acceptable distances and reddenings
(\mmm=12.1--12.3 and \ebv=0.37--0.43).
Stars in clusters or in binaries offer the best observational possibility of
estimating the mass range where AGB evolution leads to the formation of carbon
stars. In the case of NGC2660, all three best evolutionary models, even if
they give different ages for the cluster, agree remarkably well on the mass
presently at the TO: 2.0 \msun (FST) and 2.1 \msun (Padova and FRANEC).
This represents a sharp lower limit to the mass of the carbon star, and is
higher e.g., that the 1.8--1.9 \msun ~derived by Groenewegen et al. (1995).

\begin{figure}
\vspace{13cm}
\includegraphics{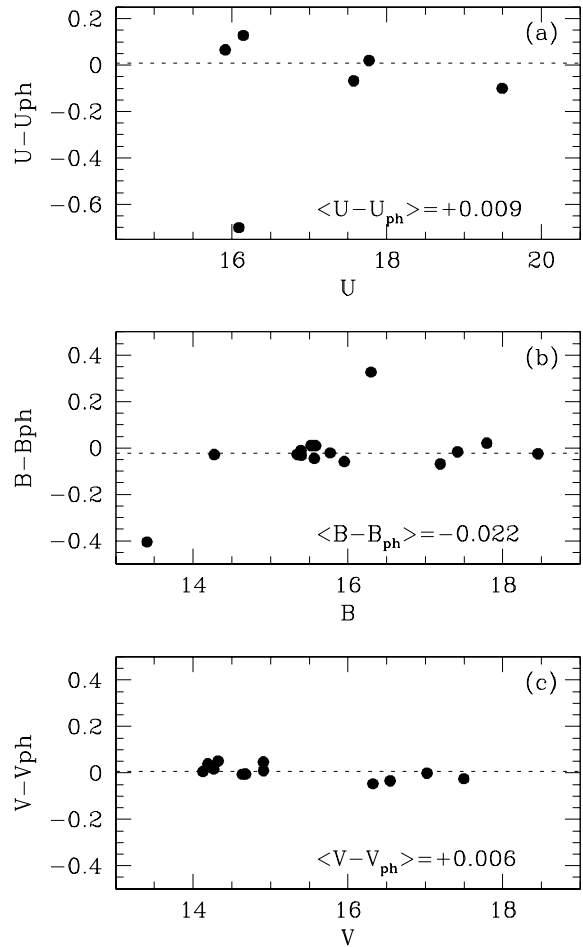}
\caption{Comparison between our calibrated magnitudes and the photoelectric
ones in HH73; also given are the mean differences (the zero points), 
computed excluding the more deviant measures (1 in U, 2 in B).}
\label{fig-conf}
\end{figure}

\subsection{Comparison with older data}

\begin{figure*}
\vspace{17cm}
\includegraphics{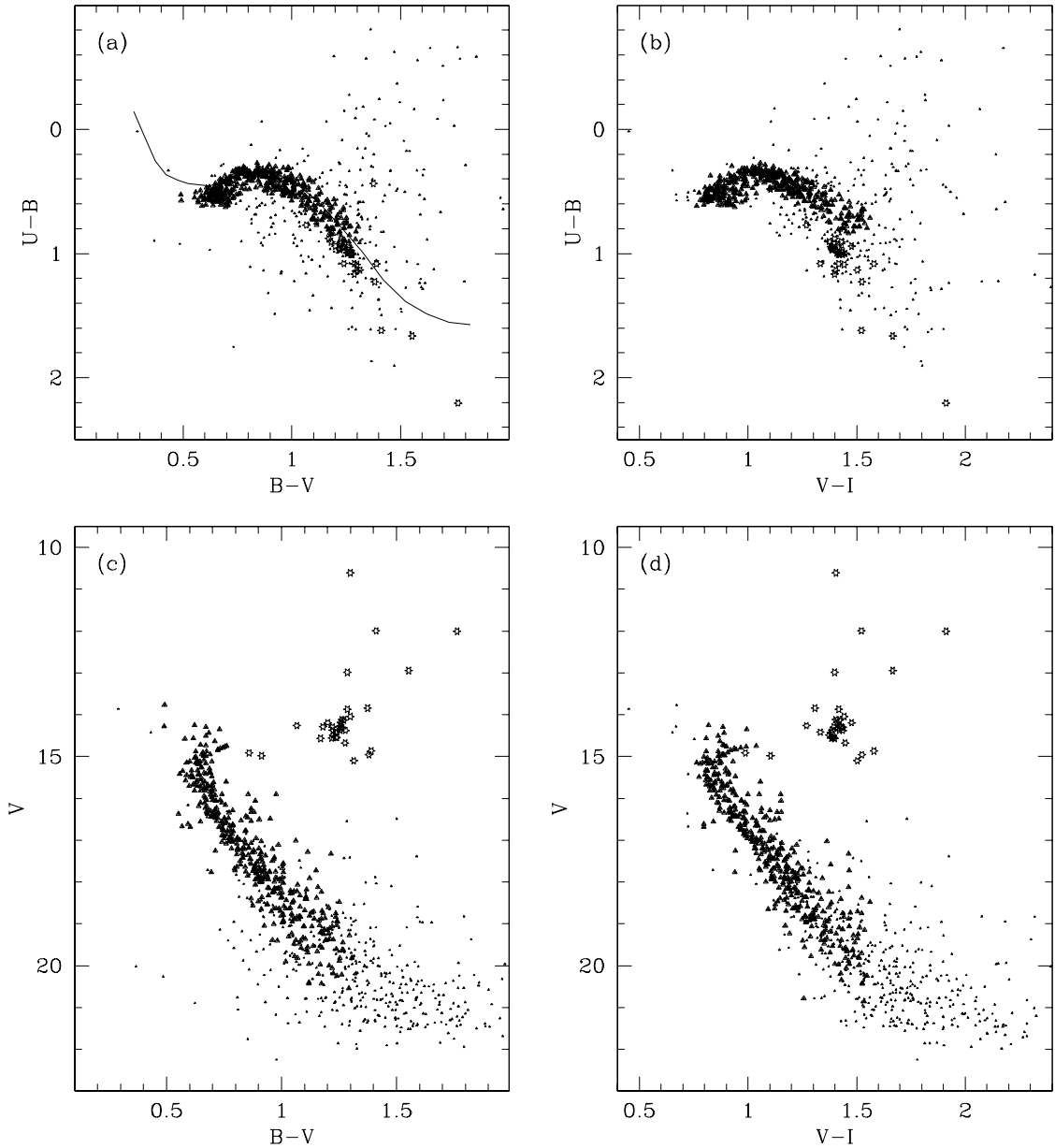}
\caption{Selection of MS stars in the two-colours diagrams; in all
panels filled triangles represent stars selected as cluster members on the MS,
small points stand for stars rejected as members, and open stars represent the
RGB/clump objects. (a) $V-B,U-B$ selection, with the ZAMS 
reddened by \ebv=0.45; (b) $V-I,U-B$ selection. (c) and (d) show which
stars were selected in the three groups in the $V,B-V$ and $V,V-I$ CMDs
respectively.}
\label{fig-trecol}
\end{figure*}

We do not attempt any comparison with FDK: our data are not suitable for a
variability search, since all frames were acquired in a short length of
time (about 2.5 hours in the same night), and we have no better information
on membership than the star positions on the CMD.
We have counteridentified 11 stars, the three for which FDK show a
light curve (their fig. 17), and a few more with parameter values 
indicating a variability larger than average. Their position on the CMD is
as follows: N$_{\rm FDK}$=160 (N$_{our}$=589), the spotted star, lies just
above the MS, where binaries are found; N$_{\rm FDK}$=168 (N$_{our}$=392)
is possibly a field star; N$_{\rm FDK}$=459 (N$_{our}$=518) is at the MS TO;
six of the others (178/427, 381/311, 461/395, 555/389, 663/409, 717/205) are
on the MS, one (417/536) is on the RGB, one (576/574) is in the Blue
Stragglers region.

A more significant test is feasible with the HH73 data. We show in
Fig.~\ref{fig-cmd}(d) their photographic $V,B-V$ diagram next to ours
(Fig.~\ref{fig-cmd}(c)): our photometry clearly shows a smaller scatter and
reaches a fainter magnitude limit.

HH73 used a set of photoelectric measurements to calibrate their data; we 
identified 15 of these stars in our field and compared our resulting
magnitudes to their photoelectric values (Fig.~\ref{fig-conf}), to test
the validity of our calibration for the {\it U, B, V} filters. 
We found small offsets; the mean differences, once we removed the extreme
deviant cases (possible misidentifications or image blends in our or in their 
photometry), are (our values minus HH73): 0.009 mag in $U$, --0.022 in $B$ 
(the frame with the worst seeing and airmass), and 0.006 in $V$.

We decided to shift our values by those very small quantities, on the basis
of the general superior quality of photoelectric magnitudes and of the
intrinsic uncertainties on the extinction correction (the errors attached to
the extinction coefficients translate to uncertainties of a few hundreths of
magnitudes). Furthermore, when comparing the Zero Age Main Sequence
(ZAMS) taken from Lang (1991) to our data
in the two-colour diagram (see Fig.~\ref{fig-trecol}(a), we obtained a good
fit with a reasonable value for the reddening \ebv=0.45 (and E($U-B$)=0.72 
$\times$ E(B--V), Cardelli, Clayton \& Mathis 1984) only with the shifted
data. This value for the reddening will be later compared to what is found 
from our CMD simulations. The comparison with this ZAMS cannot 
yield by itself the final value, since the cluster has an age large enough
(see Sect. 4) to display evolutionary effects; furthermore, we would be
assuming solar metallicity. Even if this seems to be a good enough 
approximation for NGC2660, 
the locus of populations of different metallicity in 
the two-colour diagram has not a simple dependence on metallicity (see e.g. 
Bressan, Granato \& Silva 1998, their fig.8). The I values were of course 
left unaltered; in any case, the I band is the least
affected by extinction.

The table with the photometry will be available through the CDS and BDA
(Mermilliod 1995, maintained at {\tt http://obswww.unige.ch/webda}).

\begin{figure*}
\vspace{9cm}
\includegraphics{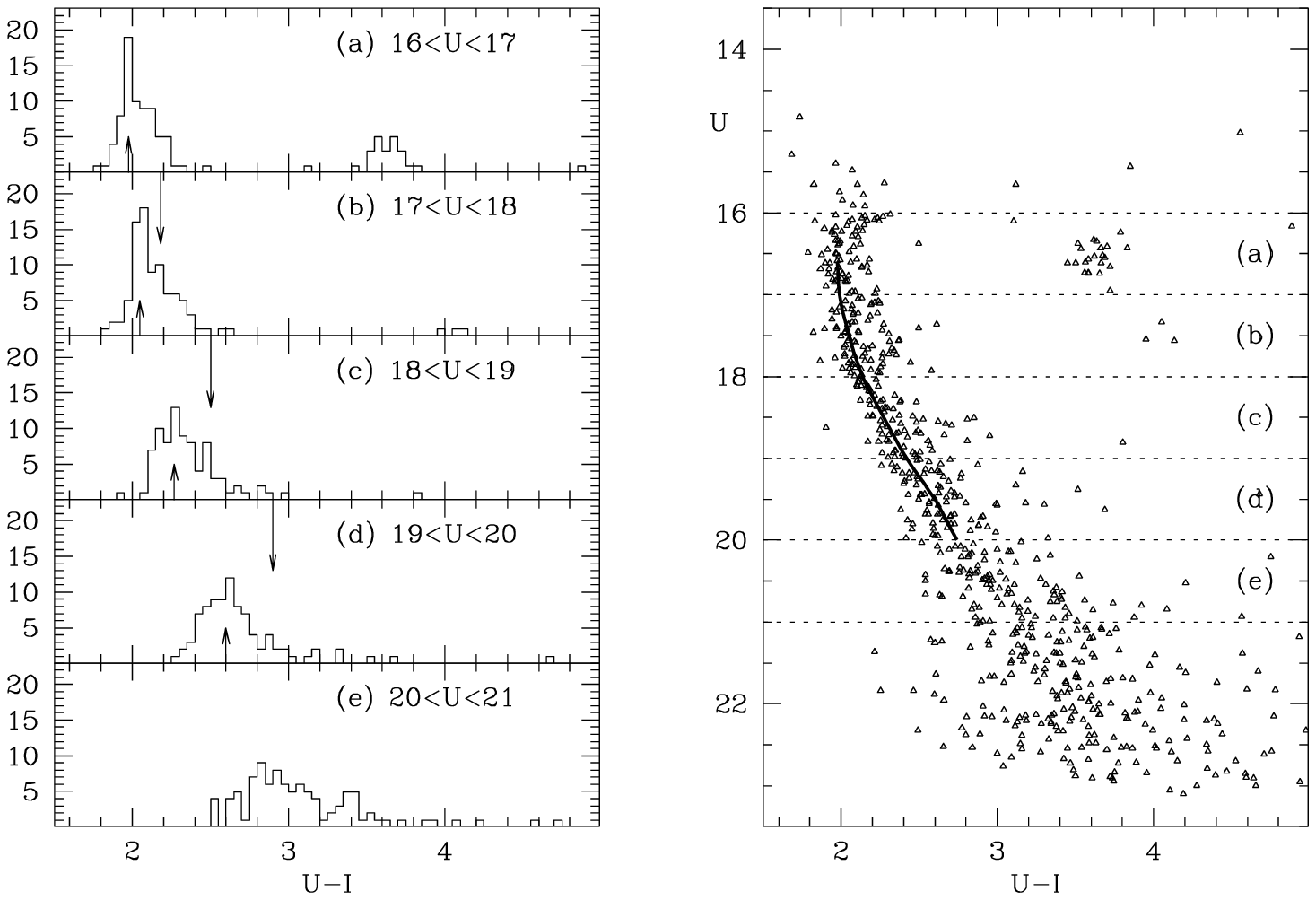}
\caption{Histogram in colour at different magnitude levels along the cluster
MS (left); the two arrows indicate the MS and the secondary peak. CMD of the
cluster in $U,U-I$, with the adopted MS ridge line (right).}
\label{fig-bin}
\end{figure*}

\subsection{Cluster members and binary sequence}

We do not have any nearby external frame to estimate the degree of
contamination by intervening field stars, but it doesn't appear very
important, especially in the red giant and clump region.

A possibility to discriminate between cluster members and field
interlopers is given by the two-colours diagrams. Given the age of the
cluster (around 1 Gyr), we are able to sample on the MS the spectral types
$\sim$ A7--K2 and fit our data to the ZAMS (Lang,
1991). We found a good fit using a reddening of \ebv=0.45. We tried to
isolate true MS members of the cluster selecting only stars falling on/near
the locus defined by the MS stars in the two-colours diagrams, both in
$B-V,U-B$ and $V-I,U-B$  (see Fig.\ref{fig-trecol}(a) and (b)). This
could look like a somewhat arbitrary process, especially at fainter
magnitudes, but it produces a reasonable CMD:
see Fig.\ref{fig-trecol}(c), with
420 MS objects selected in the $B-V,U-B$ plane, and (d), with 399 objects
selected in the $V-I,U-B$ plane.
Red giants and clump objects, for a total of 34 stars,
were all retained (they are represented in these figures by the open star 
symbols).
Finally, we chose as ``true cluster members'' all the stars present in both
selections (373 MS objects), plus the 34 giants, for a total of 407. This
sample will be later compared to the simulated CMDs.

To determine an approximate fraction of binary stars we followed the same
procedure already adopted for the other examined clusters. In this case, we 
used the $U, U-I$ diagram, since it offers the larger colour baseline, and
made histograms in $U-I$ along the cluster MS (see Fig.~\ref{fig-bin}, left
panels).
The more prominent peak corresponds to the main sequence; in the brighter
magnitude interval, the second peak represents the red clump stars. Going
fainter, a smaller secondary peak appears, that we attribute to the binaries:
its position is in agreement with the colour that a sequence brighter than the
MS by 0.75 mag (i.e. an equal-mass binary sequence) would have in each mag
interval (see Fig.~\ref{fig-bin}, right panel). Counting the stars in the two
peaks, we determine an approximate fraction of 30\% of binary stars. This
fraction turns out to be in good
agreement with that found necessary for the synthetic CMDs to well
reproduce the observed one.

\begin{figure*}
\vspace{20cm}
\caption{Reference observational CMDs of NGC2660 (top panels) 
and representative synthetic CMDs (single stars only) for the Padova
sets of stellar tracks. The left panels show the $V, B-V$ diagrams and the
right panels the $V, V-I$ ones. See text for details. }
\includegraphics{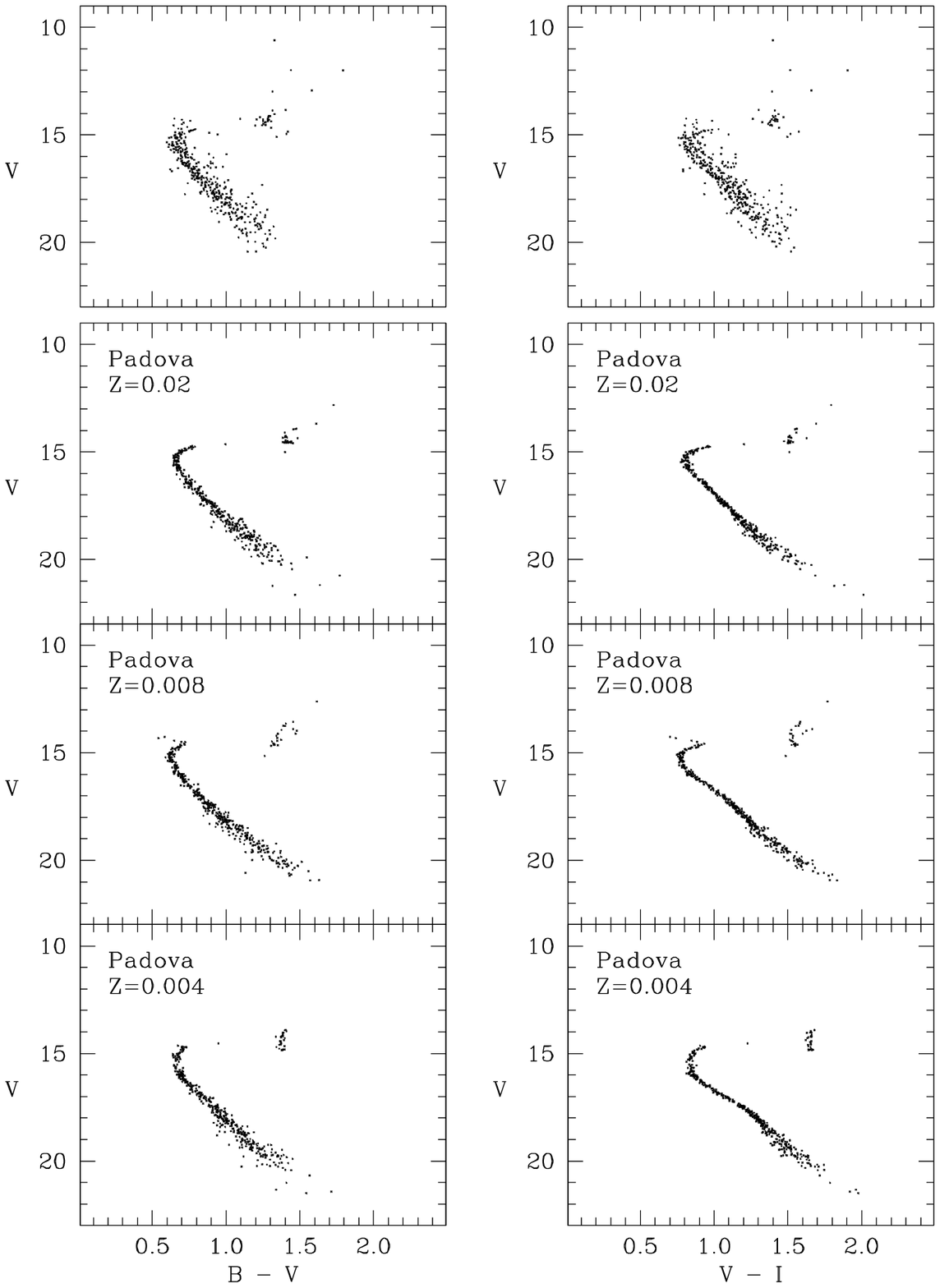}
\label{simsing}
\end{figure*}

\begin{figure*}
\vspace{20cm}
\caption{Same as Fig.~\ref{simsing} but for the FRANEC (upper panels) and FST 
(lower panels)
stellar tracks.}
\includegraphics{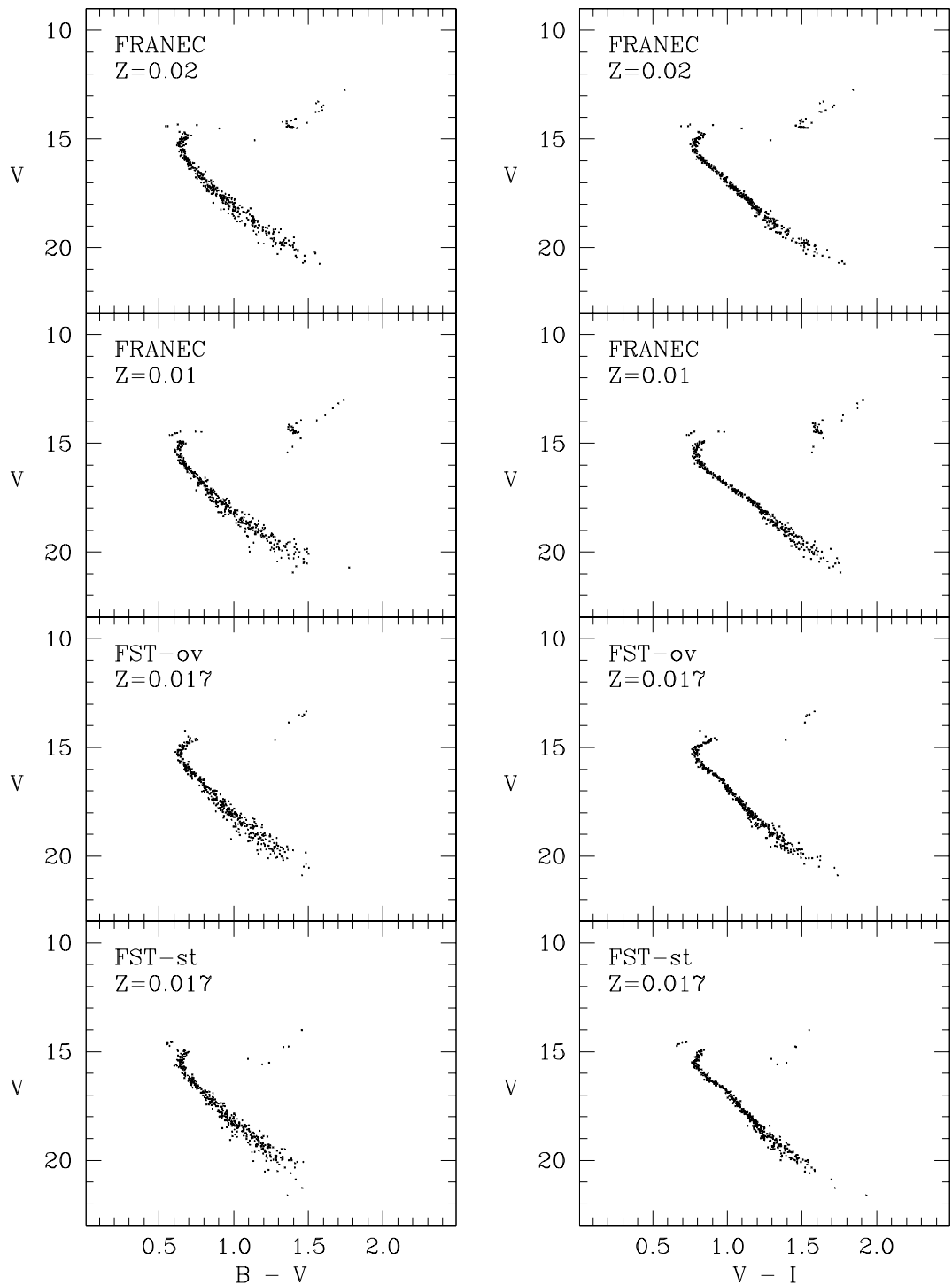}
\label{simsing2}
\end{figure*}

\begin{table*}
\begin{center}
\caption{Stellar evolutionary tracks adopted for the synthetic CMDs.}
\begin{tabular}{lccccll}
\hline\hline
Model & Y & Z & M$_{min}$    & M$_{max}$    & Reference & Notes\\
      &   &   &(M$_{\odot}$) &(M$_{\odot}$) &           & \\
\hline
FRANEC & 0.28 & 0.02 & 0.7 & 9   & Dominguez et al. 1999 & to AGB-tip\\ 
FRANEC & 0.27 & 0.01 & 0.7 & 9   & private communication & to AGB-tip\\
FST-st & 0.27 & 0.017 & 0.6 & 15 & Ventura et al. 1998  & to RGB-tip\\
FST-ov & 0.27 & 0.017 & 0.6 & 15 & Ventura et al. 1998  & to RGB-tip\\
Padova & 0.28 & 0.02 & 0.6 & 120 & Bressan et al. 1993  & to AGB-tip\\
Padova & 0.25 & 0.008 & 0.6 & 120 & Fagotto et al. 1994  & to AGB-tip\\
Padova & 0.24 & 0.004 & 0.6 & 120 & Fagotto et al. 1994  & to AGB-tip\\
\hline
\end{tabular}
\end{center}
\label{tab-mod}
\end{table*}

\section{CLUSTER PARAMETERS}

We derived the values of the cluster parameters applying to NGC2660 the
usual approach of CMD simulations described by Tosi et al. (1991) and already
employed in our previous works on open clusters. 
Due to the range of the existing 
estimates of the cluster metal content ([Fe/H] $\sim$ --1 to slightly more
than solar), the available stellar evolutionary tracks appropriate to 
create the synthetic CMDs are those
with metallicities between solar and 1/5 of solar. As in our previous 
papers, to give an estimate of the uncertainty related to the theoretical
interpretation of the CMD, we performed the simulations adopting
several homogeneous sets of stellar models. The major features of these sets 
are summarized in Table 3. The FRANEC and Padova models adopt the 
standard mixing-length theory without and with overshooting from convective 
zones, respectively. In the FST models, instead, convection is treated
following the Full Spectrum of Turbulence approach, with (FST-ov) and without
(FST-st) overshooting. For each set of stellar models,
we performed several MonteCarlo simulations for any reasonable
combination of age, reddening and distance modulus.

To avoid spurious effects due to inhomogeneous conversions from the
theoretical to the observational plane, all the model luminosities and
temperatures have been transformed to magnitudes and colours using the
same set of conversion tables, based on Bessel, Castelli \& Plez (1998 a, b)
model atmospheres. 

The incompleteness factors and the photometric errors in each magnitude bin
assigned to the synthetic stars in every photometric band are those derived
from the observed data and mentioned in section 2. The number of
stars in the synthetic diagram is 407, the same as that in the empirical CMDs 
shown in  Fig.\ref{simsing} (top panels).  The
reddening \ebv is transformed to the E(V$-$I) proper for our photometric bands
following Taylor's (1986) prescriptions.

\subsection{Results with Padova stellar models}

For these tracks we computed the synthetic CMDs with three different
metallicities: Z=0.02, Z=0.008 and Z=0.004 (Bressan et al. 1993; Fagotto
et al. 1994). Representative CMDs 
($V, B-V$ on the left and $V, V-I$ on the right) derived from each of these 
sets of models are shown in Fig.~\ref{simsing}. These
synthetic CMDs assume that all the stars in NGC2660 are single.

The solar case corresponds to a cluster age of 950 Myr, \ebv = 0.4 and
\mmm = 12.3. The Z=0.008 CMDs assume age = 1.1 Gyr, \ebv = 0.52 and 
\mmm = 12.2, and the Z=0.004 ones assume age = 1 Gyr, \ebv = 0.56 and 
\mmm = 11.9. These models were selected within each set as those with 
TO and clump stars luminosities and distributions in better agreement 
with the observed ones.

The MS gaps and the shape of MS, TO and post-MS phases are all very well
reproduced by the Z=0.02, while the Z=0.008 models show a slightly shorter
colour extension of the TO region and the metal poorer cases also show 
rounder TOs and different shapes of the MS, which worsen the fit to the
data. Besides, while the reddening required for the solar models
(0.38$\leq$\ebv$\leq$0.42) is consistent with the literature values
(0.35 to 0.38, see Introduction) and with the one found for our data in the
two-colour diagram (0.45), the reddening implied by the Z=0.008 models
(0.47$\leq$\ebv$\leq$0.52) is barely acceptable, and that for the metal poorer
models (\ebv$\geq$0.56) is in our opinion too high. We thus 
suggest that the metallicity of NGC2660 is at least half solar.

\subsection{Results with FRANEC stellar models}

The FRANEC stellar tracks have been recently updated (Dominguez et al. 1999)
to improve the input physics and include the OPAL opacities (Rogers,
Swenson \& Iglesias 1996). Thanks to 
their authors (Straniero, private communication), we have been able to use 
not only the published tracks but also others interpolated at intermediate
metallicities. Given the worse results obtained with the lower metallicity 
set of Padova models, which apply also to the FRANEC cases, we present here 
only the FRANEC results for Z=0.02 and Z=0.01.

Both the solar and the half-solar metallicity models reproduce fairly well
the detailed features visible in the observational $V,B-V$ and $V,V-I$ CMD
(like bumps and gaps on the MS), when the adopted age is between 700 and 800
Myr. The colour extension of the TO region is however slightly short and
a few subgiants are predicted where the cluster does not show them.
The results are visible in the top and second panels of 
Fig.~\ref{simsing2}, where the CMDs of single stars with Z=0.02, age=750 Myr, 
\ebv=0.42, \mmm=12.3 and those for Z=0.01, age=700 Myr, \ebv=0.53 and
\mmm=12.2 are shown, respectively. Older ages imply brighter clumps,
different shapes of the MS and too round TOs, while younger ages imply
fainter clumps and excessively hooked TOs. The same range of ages is found
to best fit the data with both metallicities. Note that these ages are
younger than those obtained with the Padova models, as a direct consequence
of the different assumption on overshooting from convective zones.

Of course, to reproduce the observed colours, the solar models require a lower
reddening than the half-solar ones, because the latter are intrinsically
bluer. In practice, the synthetic CMDs in better agreement with the empirical
one assume E$(B-V)$ around 0.5 when Z=0.01 and around 0.4 when Z=0.02. Bearing
in mind the literature value and our fit in the two-colour diagram, we
consider the above result as an indication that the metallicity of NGC2660 is
more likely solar.

No significant difference is found in the distance modulus with varying
metallicity: with Z=0.02 the best CMDs adopt \mmm=12.3, while with Z=0.01
they adopt \mmm=12.2.

\subsection{Results with FST stellar models}

The FST models (Ventura et al. 1998) are available both with and 
without overshooting from
convective zones, although only for solar metallicity. They thus allow
for a direct test, in homogeneous conditions, of the effect of overshooting
in the parameters derivation. Unfortunately, these tracks reach only the
tip of the red giant branch and have no core helium burning phases. Hence
they do not provide the extremely useful constraint represented by the
clump features. For this reason, it is more difficult to limit the range
of acceptable ages with these tracks. In fact, with the FST-st models
any age between 700 Myr and 1 Gyr seems acceptable, whereas with the FST-ov
models the range moves toward older values (0.9--1.2 Gyr) as a classical
consequence of overshooting.

Another effect of the higher luminosity related to the overshooting assumption
is the higher reddening required to reproduce the observed colours. For
the same age (and, obviously, metallicity), the FST-ov models have to assume
\ebv at least 0.03 higher than the corresponding FST-st models. However,
when the age is allowed to vary and its best range is selected, the older
age required by overshooting models makes the synthetic CMDs redder and
somewhat compensates the reddening difference. As a consequence, the
reddenings resulting in the two cases are fairly similar to each other:
overshooting models indicate \ebv between 0.37 and 0.43, and standard models
\ebv between 0.38 and 0.48.

The most striking characteristics of the FST-ov models is that the observed 
features of the MS, TO and subgiant regions are incredibly well reproduced,
even the smallest gaps or shapes which one would have not considered as 
significant. The FST-st models also reproduce well the MS features but to
a less extent. The quality of the fit can be appreciated in the bottom and
second from bottom panels of Fig.~\ref{simsing2}, where we 
show the synthetic CMDs for single stars corresponding to the FST-ov case 
with age 1.1 Gyr, \ebv=0.37 and \mmm=12.1 and to the FST-st case with age 
900 Myr, \ebv=0.40 and \mmm=12.3.

\begin{figure*}
\vspace{10cm}
\caption{Best synthetic CMDs with 30$\%$ of binary stars for each group of 
stellar models. The top panels show the $V, B-V$ diagrams and the bottom
panels the corresponding LF (solid line) overimposed on the empirical one
(dots). See text for further details. }
\includegraphics{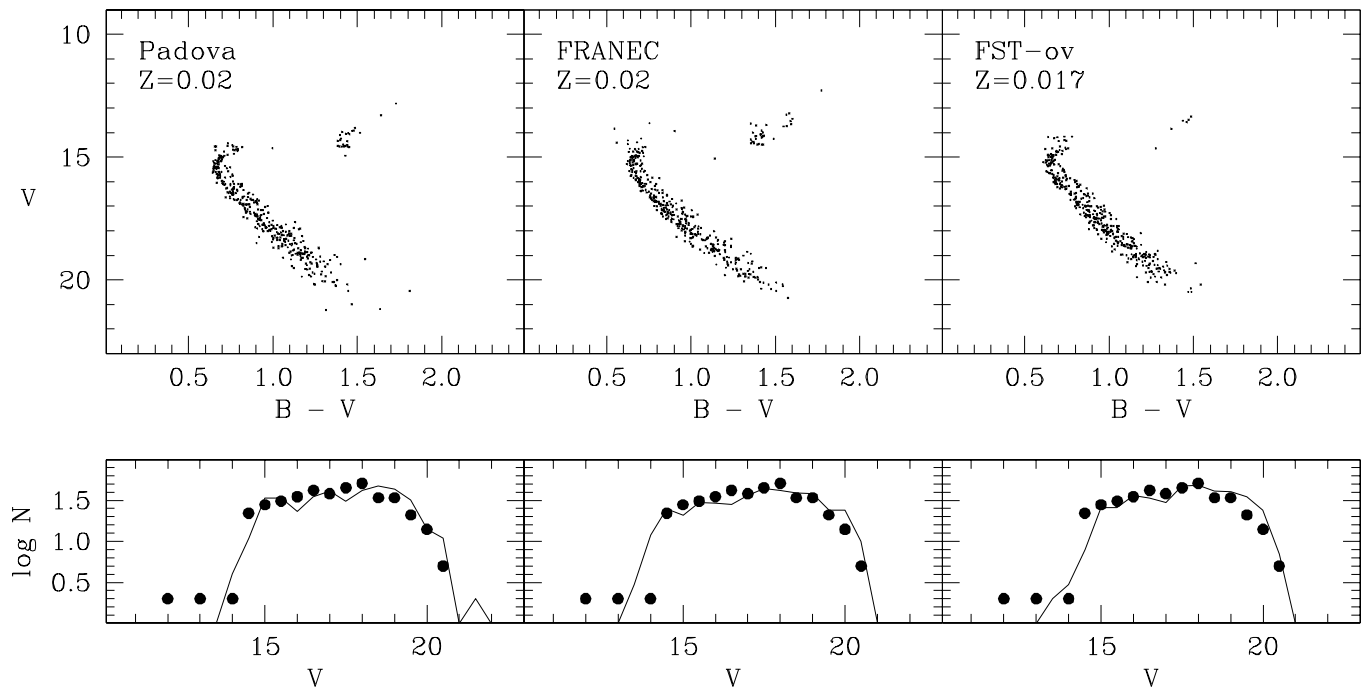}
\label{simbin}
\end{figure*}

\subsection{The need for binary stars.}

All the models in Figs~\ref{simsing} and ~\ref{simsing2} 
share the characteristics of presenting
MS bands much tighter that the empirical one. One might think that this
inconsistency comes from having underestimated the photometric errors, but
this cannot be the case, at least for the brightest magnitude bin of the TO 
region, where the photometry is very accurate, the uncertainties very small,
but the spread difference between data and predictions possibly larger
than at fainter magnitudes. The obvious explanation for such difference 
resides in the presence of a significant fraction of binary stars. 

In all the clusters already examined by our group (NGC2243, NGC2506,
NGC6253, NGC7790, Be21 and Cr261) we found evidence for a significant
fraction (30$\%$ at least) of binary stars. In NGC2660 the observational
evidence is possibly less compelling, as discussed in Section 3.2, but
the comparison between empirical and synthetic CMDs shows that binaries
must be present in this cluster as well.
Fig.~\ref{simbin} shows some of the best synthetic $V, B-V$ CMDs resulting
from the various sets of stellar evolutionary models, together with their
luminosity function (LF) overimposed to the empirical one. The only
difference with respect to the models presented in Figs~\ref{simsing} and
~\ref{simsing2} is that 136 of the 407 stars are now supposed to be in
binary systems with random mass ratios; their magnitudes and colours are
thus modified according to Maeder's (1974) prescriptions for systems of any
mass ratio between primary and secondary star.

It is apparent from Fig.~\ref{simbin} that the inclusion of this 30$\%$ of
binaries allows us to perfectly reproduce the observed spread of the
stellar distribution in all the CMD sequences, as well as the single
morphological features of the distributions (gaps, hooks, bumps, etc.). The 
observational luminosity function is also quite well reproduced by these 
models, except in the FST cases, where the lack of the clump obviously
reflects in a paucity of bright stars.

\begin{table*}
\begin{center}
\caption{Summary of best-fitting results for distance modulus, age, and 
reddening, chosen from all the simulations.}
\begin{tabular}{lrcccr}
\hline\hline
\multicolumn{1}{c}{Model} &\multicolumn{1}{c}{Z}
&\multicolumn{1}{c}{(m-M)$_0$}
&\multicolumn{1}{c}{$\tau$ (Gyr)} &\multicolumn{1}{c}{\ebv} 
&\multicolumn{1}{c}{Notes}\\
\hline
Padova & 0.02  & 12.3 & 0.95 & 0.40 & \\
FRANEC & 0.02  & 12.3 & 0.75 & 0.42 & \\
FST-ov & 0.017 & 12.1 & 1.10 & 0.37 & ages $\geq$ 0.9 Gyr acceptable\\
Padova & 0.02  & 12.3 & 1.00 & 0.38 & \\
FST-ov & 0.017 & 12.1 & 1.20 & 0.37 & ages $\geq$ 0.9 Gyr acceptable\\
FRANEC & 0.01  & 12.2 & 0.70 & 0.55 & \\
\hline
\end{tabular}
\end{center}
\label{tab-ris}
\end{table*}

\section{Discussion and Conclusions}

All the models presented above turn out to show RGB colours redder than
observed. We do not have a straight explanation for this problem, but
we envisage three possible reasons: it can be related to a) an intrinsically
too cool temperature of the stellar models, b) an inappropriate 
temperature--colour transformation in the coolest phases, c) an inappropriate
application of the reddening law. The former two options cannot be excluded,
but sound rather unlikely, since the problem equally affects all the stellar
tracks examined so far and is found to be the same with any of the 
photometric conversions applied by our group. The latter explanation seems
instead quite possible, because we applied to all the synthetic stars
a constant \ebv  (i.e. independent of spectral type), whereas Twarog, 
Ashman \& Anthony-Twarog (1997) have recently emphasized that stars of 
the RGB types show \ebv lower by 5--10 $\%$ than that of MS stars with 
intrinsic \bv ~1.0 mag bluer.
Fernie (1963) derived from a sample of supergiants the relation 
\ebv$_{true}$=\ebv $\times [0.97 - 0.09\times$(\bv)], which was
later found by Hartwick \& Hesser (1972) to hold also for red giants in
open clusters. We have applied this relation to our synthetic CMDs and found 
that it does reduce, but not sufficiently, the excessive redness of our
RGB stars. Since we have found no recent quantitative confirmation of 
the above relation, we have preferred to avoid introducing
further uncertainties. Hence, all the synthetic diagrams shown here have
been treated with a constant reddening.

We determined for NGC2660 a confidence interval for metallicity, distance,
reddening and age: metallicity about solar, \mmm = 12.1--12.3, \ebv =
0.37--0.42, age $\lsim$ 1 Gyr, with a fraction of binaries of about 30 \%. 

The cases resulting in better agreement 
with the data are listed in Table 4 where they are ranked according to the fit
quality. The ranking is the result of a quantitative comparison of
the predicted LFs with the corresponding data and of independent selections
by eye of the CMDs made by each of the authors. Such ranking is in
agreement with what can be derived from $\chi^2$ or Kolmogorov-Smirnov
tests. Nonetheless, given the uncertainties still affecting the synthetic 
CMDs for the problems mantioned at the beginning of this section, statistical
tests cannot help much in discriminating among the models and we believe that 
a better selection can still be achieved by eye, which allows to
take such uncertainties more easily into account.

We recall, once again, that the FST models have generally lower
ranks because they lack the He-burning phases (i.e. the clump) and despite 
their better fit in the computed phases.
Given the method used, the parameters attributed to the cluster are strictly
tied, and changing one of them implies changing them all. The quality 
of the fits, though, is so good, that each parameter has only a very limited 
possible range.

All the best fits are obtained for solar metallicity; furthermore, the high
reddenings implied by the more metal poor tracks are in contradiction both
with what we find from our two-colours diagram, and with literature values
based on different methods.
Our best estimate of the metallicity, based on our photometry, is in
reasonable agreement with HH73, and Lyng{\aa} (1987). The vastly different
values found in literature cannot be reconciled: the best way out of the
problem is, of course, through high resolution spectroscopic analysis.

The distance moduli resulting from 
the synthetic diagrams vary between 11.9 and 12.5,
but the extremes produce significantly worse fits, and for all the good
reproductions of the observed CMD, the distance modulus values cluster around
12.2.

The age seems to be the parameter less well determined, but the difference
is mostly between models with and without overshooting: the FRANEC
evolutionary tracks give ages consistently lower than the Padova or FST-ov
ones.
About the latter, the lack of the red clump phase implies that the age can not
be very well constrained: in fact, both the 900 Myr and the 1 Gyr models match
the observed CMD as well as the 1.1 and 1.2 Gyr ones, without significant
variations of all other parameters.

Finally, we want to stress the fact that, when all the other parameters are
the same,  models including overshooting from
the convective zones seem to better reproduce the observed data. We were
able to test it by direct comparison of models computed by the same authors
(Ventura et al. 1998), using exactly the same inputs and codes, except for
the presence/absence of overshooting.
The validity of this important test would be greatly enhanced by the 
availability of a set of tracks complete of all evolutionay phases.

\bigskip\bigskip\noindent
ACKNOWLEDGEMENTS

We warmly thank P. Montegriffo, for his expert assistance with data
reduction. We are grateful to A. Chieffi, M. Limongi and O. Straniero for
the new unpublished FRANEC tracks and for the table for photometric
conversions. P. Ventura has also kindly provided the FST tracks in
appropriate format. The bulk of the numerical code for CMD simulations has
been provided by Laura Greggio. S.S. has been partially funded on the ASI
contract ARS-96-70. This research has made use of the Simbad database,
operated at CDS, Strasbourg, France and of the BDA database, maintained by
J.C. Mermilliod. We thank J.C. Mermilliod also for his careful referee
report, which has allowed us to significantly improve the paper presentation.

\end{document}